\documentclass[naturemag,twocolumn]{revtex4-1}
\usepackage{graphicx}
\usepackage{color}
\usepackage{hyperref}
\usepackage{soul}
\usepackage{array}% http://ctan.org/pkg/array
\usepackage{fancyhdr}

\begin{document}
\newcommand{\DEOS}{DEOS}
\newcommand{\factor}{\beta}
\newcommand{\phiin}{\Delta\phi_{in}}

\newcommand{\phlam}{$^1$}
\newcommand{\desy}{$^2$}
\newcommand{\ucla}{$^3$}
\newcommand{\corresp}{$^*$}
\newcommand{\linktoarticle}{\textcolor{blue}{\href{https://www.nature.com/articles/s41377-021-00696-2}{The Version of Record of this
article is published in Light Sci. Appl. 11, 14 (2022), and is available online at https://www.nature.com/articles/s41377-021-00696-2}}}

\pagestyle{fancy}
\lhead{\linktoarticle}

\title{Phase Diversity Electro-optic Sampling:\\A new approach to single-shot terahertz waveform recording
% \\
% Phase Diversity Electrooptic Sampling: A new approach for recording terahertz waveforms in single shot\\
% Single-shot terahertz recorders using phase diversity electro-optic sampling
} 
\author{El\'eonore Roussel\phlam, Christophe Szwaj\phlam, Cl\'ement Evain\phlam, Bernd Steffen\desy, Christopher Gerth\desy, Bahram Jalali\ucla, and Serge Bielawski\phlam}
\email{serge.bielawski@univ-lille.fr}
 \affiliation{
 \phlam Univ. Lille, CNRS, UMR 8523 - PhLAM - Physique des Lasers, Atomes et Mol\'ecules,  Centre
  d'\'Etude Recherches et Applications (CERLA), F-59000 Lille, France.\\
\desy DESY (Deutsches Elektronen-Synchrotron),  Notkestr. 85, D-22607 Hamburg, Germany\\
\ucla Electrical and Computer Engineering Department, University of California, Los Angeles, 420 Westwood Plaza, Los Angeles, CA 90095, USA\\
\linktoarticle} 
% \author{C. Szwaj}
% \affiliation{\affiliationphlam}
% \author{E. Roussel}
% \affiliation{\affiliationphlam}
% \author{C. Evain}
% \affiliation{\affiliationphlam}
% \author{S. Bielawski}
\date{Accepted manuscript version, December 10, 2021}
%\pacs{41.60.Ap,29.27.Bd,05.45.-a}
\maketitle
%\begin{abstract}
{\bf
 \section*{Abstract}

{
Recording electric field evolution in single-shot with THz bandwidth is needed in science including spectroscopy, plasmas, biology, chemistry, Free-Electron Lasers, accelerators, and material inspection. However, the potential application range depends on the possibility to achieve sub-picosecond resolution over a long time window, which is a largely open problem for single-shot techniques. To solve this problem, we present a new conceptual approach for the so-called spectral decoding technique, where a chirped laser pulse interacts with a THz signal in a Pockels crystal, and is analyzed using a grating optical spectrum analyzer. By borrowing mathematical concepts from photonic time stretch theory and radio-frequency communication, we deduce a novel dual-output electro-optic sampling system, for which the input THz signal can be numerically retrieved -- with unprecedented resolution -- using the so-called phase diversity technique. We show numerically and experimentally that this approach enables recording of THz waveforms in single-shot over much longer durations and/or higher bandwidth than previous spectral decoding techniques. We present and test the proposed DEOS (Diversity Electro-Optic Sampling) design for recording 1.5 THz bandwidth THz pulses, over 20 ps duration, in single-shot. Then we demonstrate the potential of DEOS in accelerator physics by recording, in two successive shots, the shape of 200 fs RMS relativistic electron bunches at European X-FEL, over 10~ps recording windows. The designs presented here can be used directly for accelerator diagnostics, characterization of THz sources, and single-shot Time-Domain Spectroscopy.
}}
%\end{abstract}

%\begin{document}

\maketitle

\newpage
\section*{Introduction}
{Recording the complete electric field of light in single-shot (including its envelope and carrier) is considered one of the "holy grails" of terahertz science. This type of detection is largely needed for investigating and mastering novel terahertz  sources, as ultrashort pulse quantum cascade lasers~\cite{cappelli2019retrieval}, ultra-wide bandwidth laser-plasma-based terahertz sources~\cite{kress2004terahertz}, and terahertz Free-Electron-Lasers~\cite{hafez2018extremely}. Such tools are also crucial for mastering novel "extreme photonic infrastuctures", such as X-ray Free-Electron Lasers~\cite{decking2020mhz}, and in the very active field of Laser-Plasma Acceleration (LPA)~\cite{buck2011real}, which has the ambition to replace large accelerators facilities by table-top installations.}
{Single-shot terahertz recorders are also needed in applications, such as spectroscopy, using high power THz sources. In this case, the low repetition rate of the sources makes traditional methods of time-domain spectroscopy (based on scanning the delay between a probe laser and the THz signal under interest) largely impractical.}

% This includes investigations of rapidly evolving reactions in chemistry and biology~\cite{hafez2018extremely}, as well as nonlinear physics (such as nonlinear spectroscopy) which use low repetition rate high power terahertz sources. From the industrial point of view, single-shot terahertz recorders would also introduce a disruptive competition with respect to latest fast scanning terahertz spectrometers, such as Asynchronous Optical Sampling (ASOPS) and Electronically Controlled Optical Sampling (ECOPS)~\cite{bartels2007ultrafast,kim2010high}, in the quest for non-destructive and high throughput inspection for, e.g., automotive and drug industries.

\begin{figure*}[]
\begin{center}
\includegraphics[width=16cm]{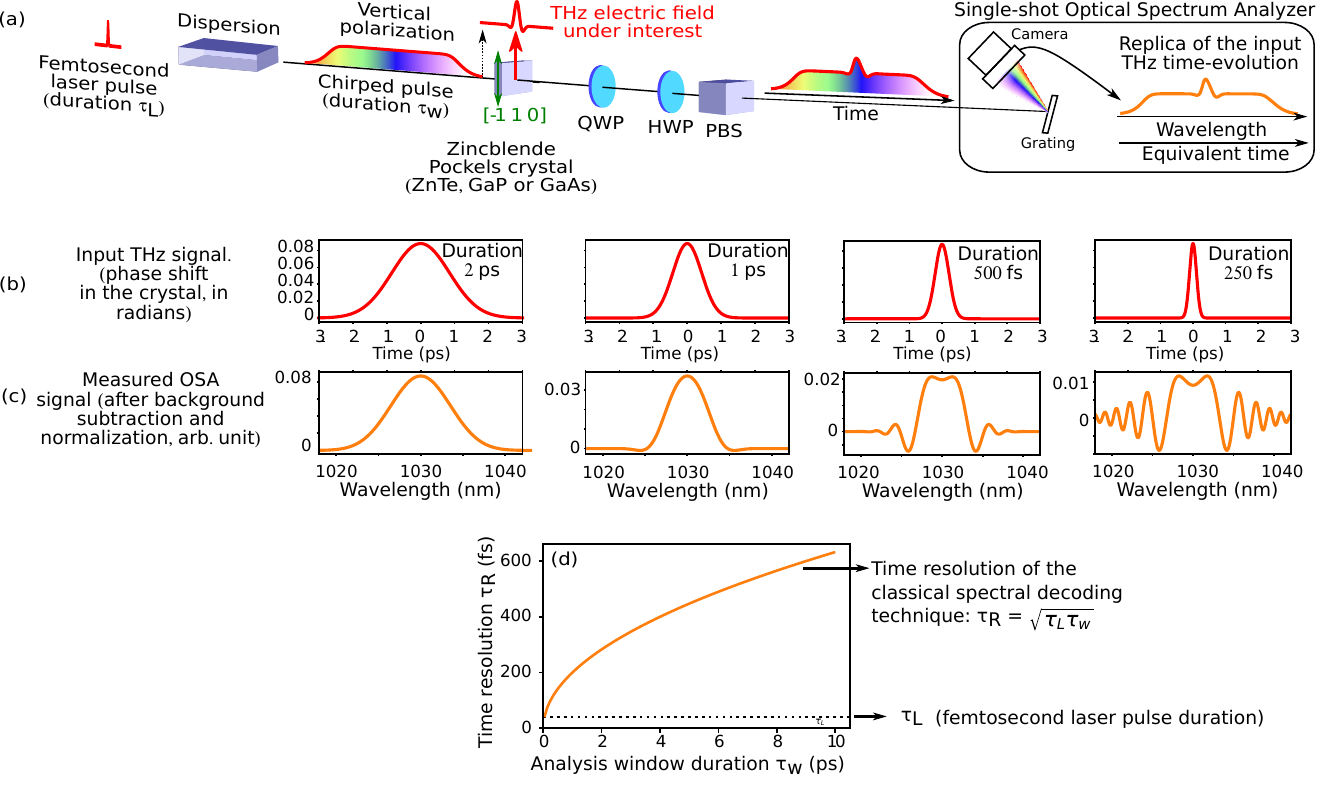}
\caption{{Principle and limitations of classical single-shot THz waveform recorders using time-to-spectrum conversion (also known as spectral decoding). {\bf (a)} Principle. The electric field pulse shape modulates the birefringence of a Pockels electro-optic crystal. A probe chirped laser pulse (with stretched duration $\tau_w$) is then intensity-modulated in single shot after passing the Pockels crystal, the quarter and half-wave plates (QWP and HWP) and the polarizing beam-splitter (PBS). Because of the laser chirp, the input temporal shape is expected to be "replicated" in the laser spectrum recorded by the grating OSA. {\bf (b)} and {\bf (c)}. Fundamental time-resolution limitation of the method (numerical simulation assuming a perfect crystal with infinite bandwidth). The method is unreliable (i.e., strong deformations occur) when the input THz pulse is shorter than $\tau_R=0.7$ ps, although a 39~fs femtosecond laser is used. (d) resolution limitation of the classical method {(orange, from Eq.~(\ref{eq:resolution_limit})). An objective of DEOS is to remove this limitation, and obtain a resolution that does no more degrades when the duration of the analysis window $\tau_w$ is increased. The probe laser compressed duration $\tau_L$ (black line) is given for reference.} Laser parameters: 1030~nm wavelength and 40~nm FWHM bandwidth (i.e., $\tau_L=39$~fs compressed laser pulse duration). $\tau_w=10$~ps FWHM. See Table~\ref{table:orientations} for crystal orientations, and Material and Methods for details).}}
\label{fig:fig0}
\end{center}
\end{figure*}

{However recording a complete terahertz wave in single-shot is a largely open problem, when a large bandwidth and long record duration are simultaneously needed. Promising strategies consist in extending the so-called electro-optic sampling technique~\cite{valdmanis1982picosecond} to the single-shot case. A popular strategy consists of imprinting the electric field evolution onto a chirped laser pulse~\cite{EOS_first_spectral_encoding,sun1998analysis,yan2000subpicosecond}, and recording the optical spectrum using a grating-based optical spectrum analyzer (OSA), as displayed in Figure~\ref{fig:fig0}a. This method is now usually known as {\it spectral encoding} or {\it spectral decoding}.}

{The initial idea~\cite{EOS_first_spectral_encoding}, has been based on the assumption that -- because the THz signal is imprinted onto the time-evolution of a chirped laser pulse -- we can expect the THz signal shape to appear, with a good fidelity, in the output optical spectrum.} {{However it was shown that this idea works only if the needed temporal resolution is larger than}~\cite{sun1998analysis}:
\begin{equation}
\tau_R=\sqrt{\tau_w\times\tau_L}\label{eq:resolution_limit},
\end{equation}
where $\tau_w$ is the duration of the chirped pulse on the crystal, and $\tau_L$ is the Fourier-transform limit duration (i.e., the pulse duration that may be reached when fully compressed).}

{It is important to note that this limitation is much more drastic than a simple "blurring effect". {When time evolutions faster than $\tau_R$ are present in the THz signal}, the output signal is typically a very deformed version of the input signal, as displayed in Fig.~\ref{fig:fig0}c. Moreover the limitation set by Equation~(\ref{eq:resolution_limit}) dramatically worsens when the desired analysis window duration increases (see Figure~\ref{fig:fig0}d). As a concrete example, in a situation where a recording duration of $10$~ps is required, and a $100$~fs laser is available, the resolution $\tau_R$ would be limited to $1$~ps. More generally this limitation implies that the resolution $\tau_R$ will be much larger than the initial pulse duration $\tau_L$, this issue being worse as the desired analysis window~$\tau_w$ increases.}

{This major issue} has been the subject of active investigations during the last two decades. Deconvolution algorithms have only led to limited improvements~\cite{yellampalle2005algorithm,wu2018electro}, and most research switched to alternate hardware strategies, such as encoding the information onto the transverse or angular direction~\cite{shan2000single,srinivasan2002novel,kawada2011single,minami2013single,kim2007single}, or combining spectrally decoded electro-optic sampling with advanced laser pulse characterization techniques (such as FROG)~\cite{jamison2003high,walsh2015time}. This led to improvements of the temporal resolution. However these complex designs introduce new trade-offs between sensitivity, maximum recording duration, temporal resolution, { and repetition rate.}

{Here we show that it is possible to considerably increase the temporal resolution of {\it spectral decoding} measurement systems by using a new approach that we call Diversity Electro-Optic Sampling (DEOS). This modification allows us to "break" the previous fundamental barrier displayed by Eq.~(\ref{eq:resolution_limit}), and reach sub-picosecond resolution without fundamental limitation on the window of analysis. Key to our approach is a completely new conceptual approach of spectrally decoded electro-optic sampling, which has its roots in the theory of the {\it photonic time-stretch analog-to-digital converter} ~\cite{time_stretch_first_bhushan1998time,fard2013photonic} where a similar problem called dispersion penalty was identified and solved ~\cite{han_2005_phase_diversity}. As we will see, this point of view provides a way to remove the temporal resolution limit by using a technique known as {\it phase diversity}~\cite{han_2005_phase_diversity}, and which makes use of a dual-output electro-optic sampling system. 
Note that name phase diversity comes from its use in time stretch systems and pays homage to antenna diversity, a technique that eliminates the transmission fading produced by multipath interference, by using several transmission channels (i.e., several antenna).  }

{We describe the {\DEOS} technique and validate it on two very different experimental applications. First we show how to achieve phase diversity utilizing polarization modulation in an electro-optic crystal and demonstrate single-shot recording of THz pulses in a table-top environment. This provides an experimental validation of the method, as well as the building blocks of a novel "single-shot Time-Domain Spectroscopy" system. Second, we demonstrate the {\DEOS} approach in an electron accelerator, by probing the {Coulomb-field} of electron bunches at megahertz repetition rates, at the European X-ray Free-Electron Laser~\cite{decking2020mhz} (EuXFEL). This experiments achieved {sub-200~fs resolution over a time window in the 10~ps range.}}

\begin{figure*}[]
\begin{center}
\includegraphics[width=16cm]{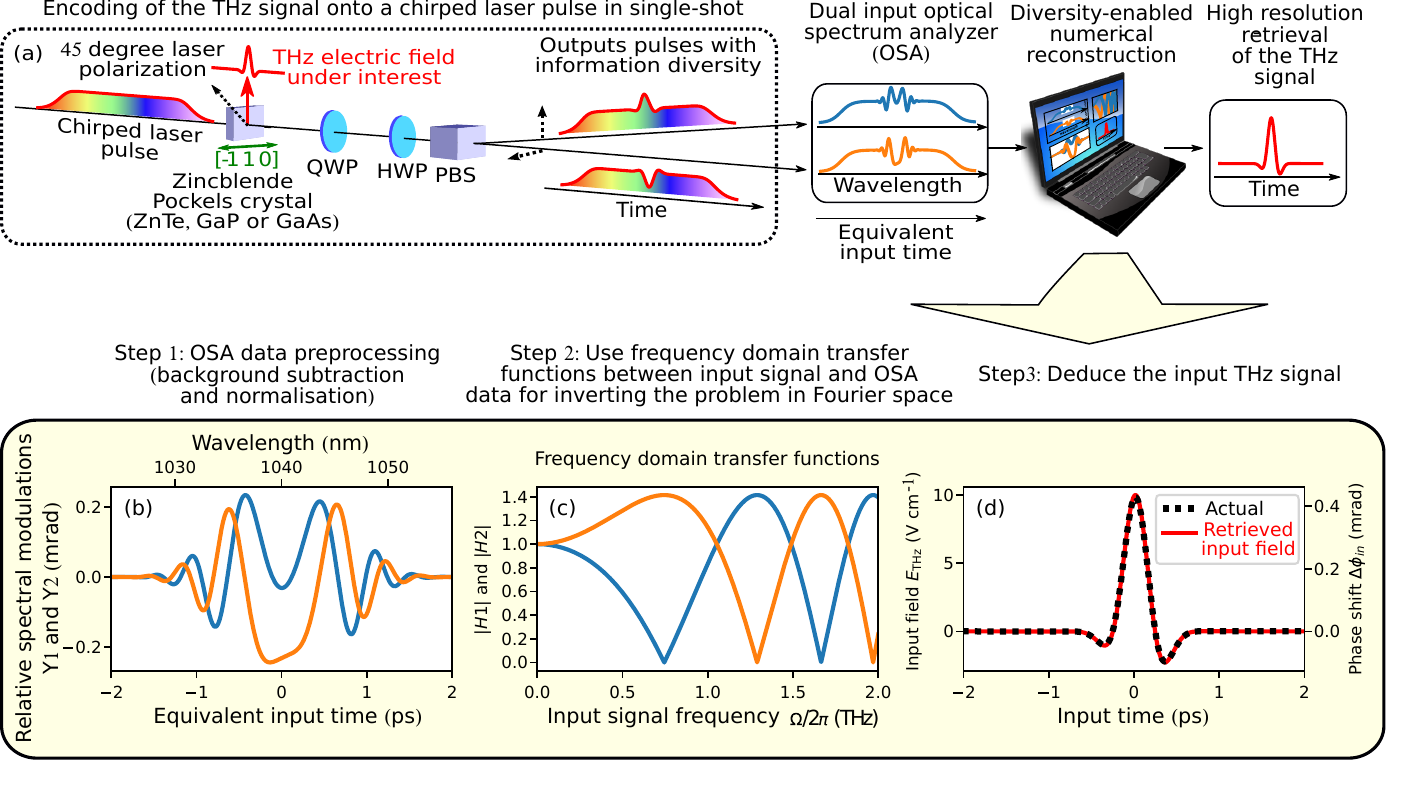}
\caption{{Principle of Phase Diversity Electro-optic Sampling, {\DEOS}, for single-shot recording of THz electric fields. {\bf (a)} Experimental design: The input THz field evolution modulates a chirped laser pulse, and the {\DEOS} design provides two outputs that contain different information. Using this two-output design (see Table~\ref{table:orientations} for details), the recorded information is sufficient for removing the problem, i.e., retrieving the input signal with high resolution. {\bf (b,c,d)} Main steps of our reconstruction method (numerical simulation). {\bf(b)} Raw electro-optic sampling (EO) signals (optical spectra after background subtraction and normalization). {\bf (c)} Transfer functions $H_1$ and $H_2$ corresponding to the two polarization outputs showing the phase diversity operation. {\bf (d)} Input signal retrieved from the recorded OSA signals [displayed in (b)] and the transfer functions $H_{1,2}(\Omega)$ [displayed in (c)], using the MRC algorithm [Eq.~(\ref{eq:MRC})]. Laser parameters close to those of the European XFEL experiment presented hereafter (1040~nm wavelength and 40~nm FWHM bandwidth, $\tau_w=5$~ps FWHM).
%Same laser parameters as for Fig.~\ref{fig:fig0} -- see Table~\ref{table:orientations} for crystal orientations.
 }}

\label{fig:fig1}
\end{center}
\end{figure*}

\section*{Results}
\subsection*{{\DEOS} single-shot recorder: experimental technique and novel theoretical framework}
Our experimental method is displayed in Figure~\ref{fig:fig1}a. {\DEOS} is based on chirped pulse electro-optic sampling, with a readout that uses a grating spectrometer~\cite{EOS_first_spectral_encoding,PhysRevLett.88.124801,sun1998analysis,peng2008optimal,fletcher2002distortion,schmidhammer2009single,PhysRevSTAB.15.070701,funkner2019high}, but with crucial modifications that remove the temporal resolution limitation. The electric field under test $E(t)$ is imprinted in single-shot onto the intensity and phase of a chirped laser pulse. If we assume a linear chirp, the optical angular frequency at the spectrometer $\omega_{opt}$ is related to the input time $t$ by:
\begin{eqnarray}
t=-\frac{\omega_{opt}-\omega^{center}_{opt}}{C}\label{eq:time_wavelength}
% \text{with}\;\lambda=2\pi c/\omega_{opt}
\end{eqnarray}
where $C=\partial\omega_{opt}/\partial t$ is the laser chirp rate. 
\par
A key innovation in {\DEOS} is how to achieve phase diversity in polarization-based electro-optic modulation. To do so, we record the optical spectra in single-shot at the two polarizer outputs, and the two optical spectra provide the input data of our retrieval algorithm. The optical spectra are first passed through a pre-processing pipeline that consists of subtraction, amplitude scaling and optical wavelength to time conversion (see Methods). These signals $Y_1(t)$ and $Y_2(t)$ will be named the EO (electro-optic) signals in what follows. Examples of EO signals are displayed in Figure~\ref{fig:fig1}b. A key feature of the {\DEOS} method that creates diversity is the special arrangement of the crystal, waveplates, and polarizations (see Table~\ref{table:orientations}), which leads to {\it diversity in frequency response} on the two output channels. This diversity will enable a retrieval of the input terahertz signal using the maximum ratio combining ~\cite{kahn1954ratio,han_2005_phase_diversity}.

{At first glance, it may not be obvious how the two output channels exiting the polarizing beam-splitter achieve phase diversity. While the two outputs may appear to carry the same information, as displayed in Table~\ref{table:orientations}, the {\it phase} modulations will be different. This {\it phase diversity} phenomenon~\cite{han_2005_phase_diversity} will thus lead to the two OSA channels to have complementary null behavior (see Figure~\ref{fig:fig1}). We will see below that the combined information in $Y_1(t)$ and $Y_2(t)$ will eliminate the nulls in the frequency response and hence overcome the fundamental bandwidth limitation}.

\begin{table*}[htbp]
\begin{center}
\begin{tabular}{ |c|c|c| } 
 \hline
  & Classic spectrally decoded EO sampling & DEOS  \\ 
  \hline
 Pockels crystal & [-110] axis {\bf parallel} to THz field & [-110] {\bf perpendicular} to THz field\\ 
 \hline
 Input laser polarization & at {\bf 0 or 90 degrees} %{\bf parallel or perpendicular} 
 wrt THz field & at {\bf 45~degrees} wrt THz field \\ 
 \hline
QWP (*) & one axis at {\bf 45~degrees} wrt [-110]& one axis along [-110]  \\ 
\hline
HWP (*) & optional & at {\bf 22.5~degrees} wrt [-110] \\ 
 \hline
 Transfer functions $H_1$ and $H_2$ & \raisebox{-0.5\totalheight}{\includegraphics[width=5cm]{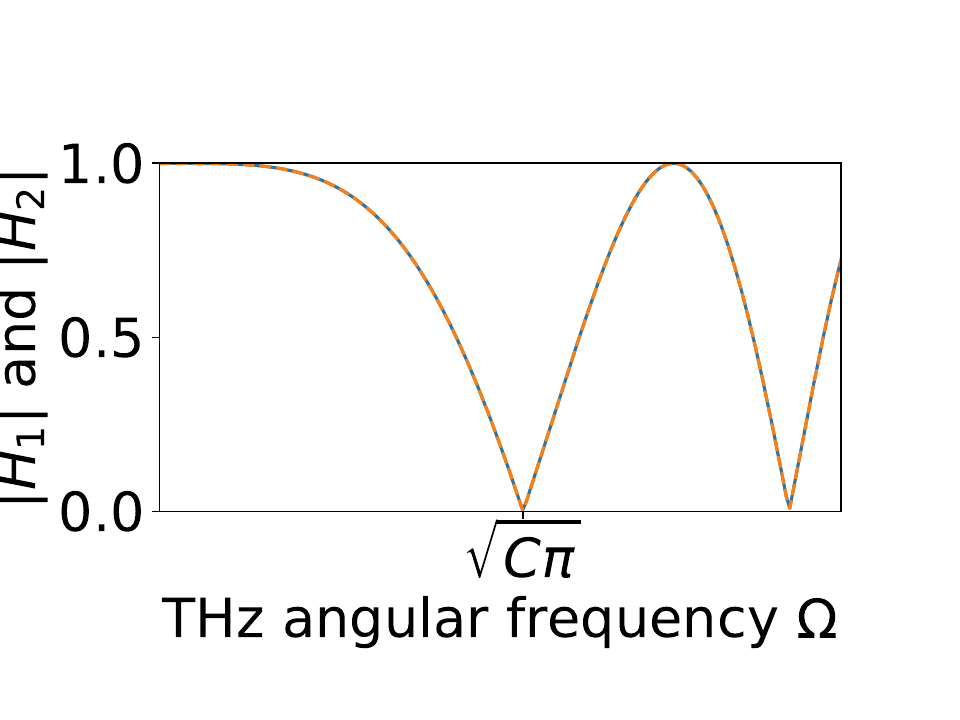}}  & \raisebox{-0.5\totalheight}{\includegraphics[width=5cm]{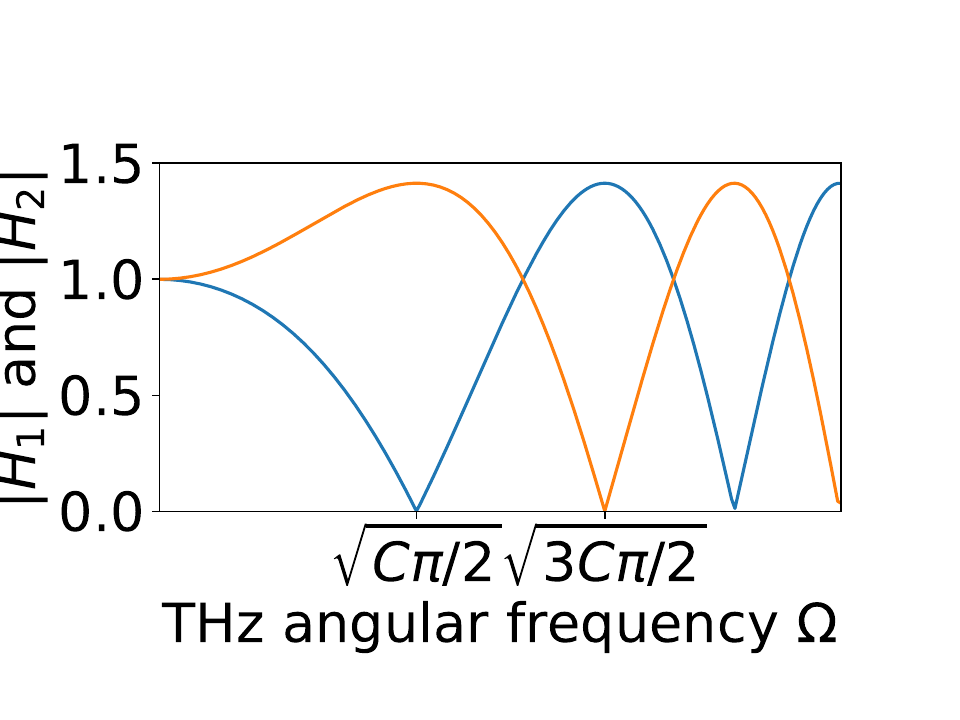}} \\
 \hline
 \end{tabular}

% \begin{table}[ht]
% \begin{center}
% \begin{tabular}{ |c|c|c| } 
%  \hline
%   & Classic spectrally decoded EO sampling & DEOS  \\ 
%   \hline
%  Pockels crystal & [-110] axis {\bf parallel} to THz field & [-110] {\bf perpendicular} to THz field\\ 
%  \hline
%  Input laser polarization & {\bf same} as THz field & at {\bf 45~degrees} wrt THz field \\ 
%  \hline
% QWP (*) & one axis at {\bf 45~degrees} wrt [-110]& one axis along [-110]  \\ 
% \hline
% HWP (*) & optional & at {\bf 22.5~degrees} wrt [-110] \\ 
%  \hline
%  Transfer functions & $H_{1}=-H_2=\cos{\frac{1}{2C}\Omega^2}$ & $H_{1}=\sqrt{2}\cos{\left(\frac{1}{2C}\Omega^2+\frac{\pi}{4}\right)}$\\
%  &&$H_{2}=-\sqrt{2}\cos{\left(\frac{1}{2C}\Omega^2-\frac{\pi}{4}\right)}$\\
%  \hline
%  First nulls of $H_1$ and $H_2$ &$\sqrt{C\pi}$ (identical) &$\sqrt{C\pi/2}$ and $\sqrt{3C\pi/2}$ (interleaved)\\
%  \hline
%  \end{tabular}

\caption{{Differences in crystal and waveplates orientations between classic spectrally decoded  EO sampling (Fig.~\ref{fig:fig0}) and the new phase diversity ({\DEOS}) system (Fig.~\ref{fig:fig1}). For the classical system, we indicate the most commonly used crystal orientation, which corresponds to maximal Pockels effect. (*): For classic spectrally decoded EO sampling, several quarter wave plates (QWP) and half wave plates (HWP) configurations are possible and we indicate here the angles for the so-called balanced detection. Note that for {\DEOS}, the orientations are chosen in order to obtain information {\it phase diversity} at the two outputs, i.e., transfer functions with interleaved zeros (see also Figure~\ref{fig:fig1}c). $\Omega$ is the input THz signal angular frequency, and $C$ is the laser chirp rate [see text and Eq.~(\ref{eq:time_wavelength})].}}

\label{table:orientations}
\end{center}
\end{table*}
\subsection*{Theoretical framework for Phase Diversity}
{In this section we derive the relation between the input $E(t)$ and the outputs $Y_1(t)$ and $Y_2(t)$ measured by the single-shot grating-based optical spectrum analyzer. After a relatively non-trivial calculation (see Supplementary Material, Sections~V-VI), it can be shown that, in Fourier domain, the input and outputs are related by remarkably simple analytic expressions involving {\it transfer functions} $H_1$ and $H_2$:}
\begin{eqnarray}
\tilde Y_1(\Omega)\approx H_1(\Omega)\tilde{\phiin}(\Omega)\label{eq:Y1H1_general}\\
\tilde Y_2(\Omega)\approx H_2(\Omega)\tilde{\phiin}(\Omega)\label{eq:Y2H2_general}
\end{eqnarray}
where the tilde denotes the Fourier transform, and $\Omega$ is the terahertz angular frequency at the input. $\phiin(t)$ is the optical phase shift induced by the THz field $E(t)$ in the crystal birefringence: 
\begin{eqnarray}
\phiin(t)&=\factor E(t)\\
\text{with}\;\factor&=\frac{\pi d}{\lambda}n_0^3r_{41}
\end{eqnarray}
{where $n_0$ is the refractive index at vanishing electric field, $d$ is the thickness of the crystal and $r_{41}$ is the electro-optic coefficient}. $\lambda$ is the laser wavelength in vacuum and $E(t)$ the electric field inside the crystal. The transfer function approach is valid for small values of $\phiin(t)$.   
% Note that the value of $\factor$ is two times smaller than for the usual design used in classical balanced detection. The transfer functions $H_{1,2}(\Omega)$ depend on the  orientation of the s, crystal, and laser polarization.

For the specific {\DEOS} design described in Figure~\ref{fig:fig1}a and Table~\ref{table:orientations}, the corresponding transfer functions can be written {as} (see Supplementary Material, Section~VI):
\begin{eqnarray}
&&H_1(\Omega)=\sqrt{2}\cos{\left(B\Omega^2+\frac{\pi}{4}\right)}\label{eq:H1}\\
&&H_2(\Omega)=-\sqrt{2}\cos{\left(B\Omega^2-\frac{\pi}{4}\right)}\label{eq:H2}
\end{eqnarray}
where $B=\frac{1}{2C}$. 

As a first observation, the transfer functions $H_1(\Omega)$ and $H_2(\Omega)$ present nulls at specific frequencies (Fig.~\ref{fig:fig1}c). These nulls limit the temporal resolution and the loss of signal at or near the null frequencies cannot be reversed by deconvolution (i.e., to retrieve the input field $E(t)$ from the recorded data). This theoretical result on the existence of nulls is consistent with previous experimental observations~\cite{murakami2008dependence,van2008single}, and ours, as we will see below. {Conceptually, these nulls correspond to the dispersion penalty phenomenon observed in the photonic time-stretch digitizer~\cite{han_2005_phase_diversity}, as shown in Supplementary Material, Sections~VB and~VIA.}

With the deep insight gained from the theoretical analysis above, we can recover the input signal $\phiin(t)$ and $E(t)$ by exploiting the information contained in both output channels because $H_1(\Omega)$ and $H_2(\Omega)$ are complementary (diverse). In this respect, it is important to note that the relative positions of the zeros (nulls) depend on the crystal orientations. {For instance, the polarization orientations used traditionally ([-110] axis parallel to the terahertz field {-- see Table~\ref{table:orientations} and in Fig.~\ref{fig:fig0})} would not work, as this would lead to} nulls of $H_{1,2}$ that occur at the same frequencies (see Supplementary Figure S4). In addition, the polarization arrangement considered in this article -- for which the zeros are complementary -- is only one of many possible choices which are compatible with the following retrieval algorithm.

%\bigskip
%******

\subsection*{Maximal Ratio Combining (MRC) algorithm for signal reconstruction}
As the zeros (nulls) of the transfer functions are interleaved, we can retrieve the unknown THz electric field from the recorded data. The mathematical problem is well-posed, and even overdetermined, i.e., one has the freedom to chose either $H_1$ or $H_2$ for inverting the problem, except at the nulls for which both channels must be used. Several advanced methods exist for the reconstruction process. Here we use the so-called Maximal Ratio Combining technique (MRC)~\cite{kahn1954ratio,han_2005_phase_diversity}, which is designed for optimizing the signal-to-noise ratio (SNR). The input signal can be retrieved from the measurements $Y_{1,2}$ as~\cite{han_2005_phase_diversity}:
\begin{eqnarray}
&&\tilde E_{in}^{retr}(\Omega)=\frac{1}{\beta}\tilde{\phiin^{retr}}(\Omega)\\
&&\text{with}\,\tilde{\phiin^{retr}}(\Omega)=\frac{H_1(\Omega)\tilde Y_1(\Omega)+H_2(\Omega)\tilde Y_2(\Omega)}{H_1^2(\Omega)+H_2^2(\Omega)}\label{eq:MRC}
\end{eqnarray}
where $\tilde E^{retr}_{in}(\Omega)$ and $\tilde{\phiin^{retr}}(\Omega)$ are the retrieved input electric field and crystal phase modulation, expressed in Fourier space. The input signal $E^{retr}_{in}(t)$ (or equivalently ${\phiin^{retr}}(t)$) is then obtained by performing an inverse Fourier transform.

 {Note that the theoretical transfer function of DEOS, defined from the input THz-induced crystal phase-shift $\Delta\tilde\phi(\Omega)$ to the reconstructed signal $\Delta\tilde\phi_{in}^{retr}(\Omega)$ is -- by design -- perfectly flat. This can be easily verified by injecting the definitions of the single channel transfer functions [Eqns.~(\ref{eq:Y1H1_general},\ref{eq:Y2H2_general})] in the MRC reconstruction Equation~(\ref{eq:MRC}). One easily finds:}
\begin{eqnarray}
&&\Delta\tilde\phi_{in}^{retr}(\Omega)=H_{MRC}(\Omega)\Delta\tilde\phi_{in}(\Omega),\\
&&\text{with}\; H_{MRC}(\Omega)=1,\label{eq:HMRC}
\end{eqnarray}
{or, after an inverse Fourier transform:}
 \begin{eqnarray}
\Delta\phi_{in}^{retr}(t)=\Delta\phi_{in}(t).\label{eq:responseMRC}
\end{eqnarray}
{This implies that any sharp input temporal feature $\Delta\phi_{in}(t)$ with bandwidth $\Omega_{max}$ should now be retrieved without distortion, provided the corresponding input data $\tilde Y_1(\Omega)$ and $Y_2(\Omega)$ can be obtained with reasonable signal-to-noise ratio, up to $\Omega_{max}$.}

We tested this reconstruction method numerically for various parameters, and found that it is possible to retrieve the input pulse for arbitrarily long input chirped pulses (i.e., for any duration of the analysis window), down {to terahertz pulse durations of the order of the Fourier-limited pulse duration (see Discussion, Suplementary Material, Section~VIII)}. An example of retrieved input pulse is presented in Figure~\ref{fig:fig1}d.

\subsection*{Experimental demonstration: recording free-propagating terahertz pulses with large time-bandwidth products}
\begin{figure*}[]
\begin{center}
\includegraphics[width=16cm]{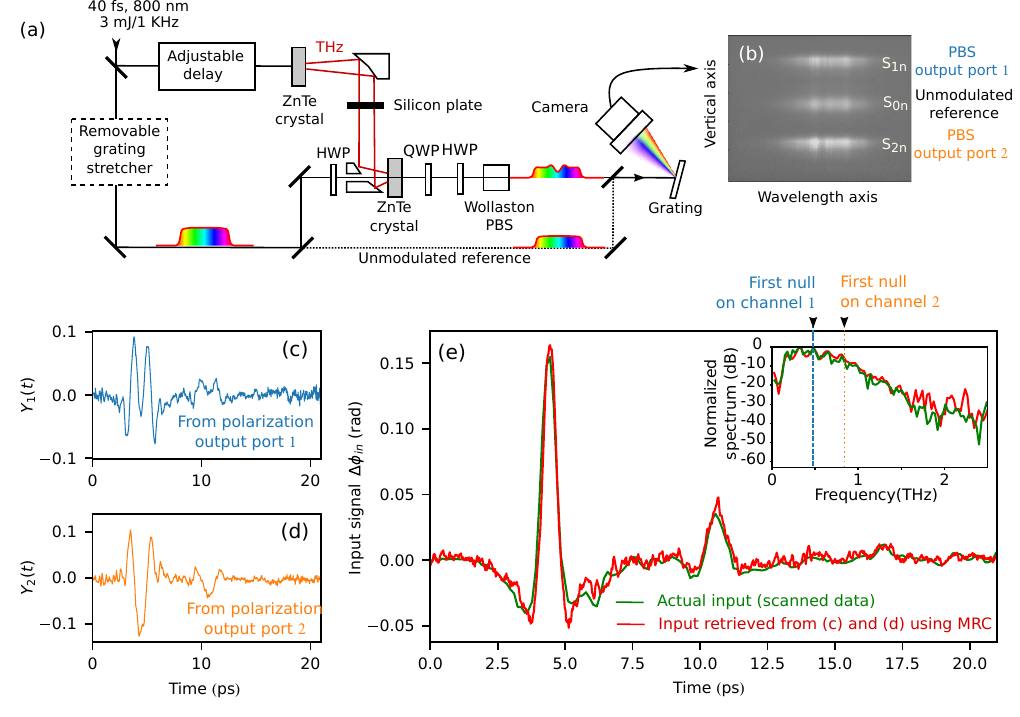}
\end{center}
\caption{Single-shot recording of free-propagating terahertz pulses over a window of the order of 20~ps. {\bf (a)}: {\DEOS} experimental setup. ZnTe: 1~mm-thick, 110-cut Zinc Telluride crystal, HWP: Half-wave plates, QWP: Quarter-wave plate, PBS: Wollaston polarizing cube beam-splitter. The beams emerging from the PBS are in the plane perpendicular to the figure. {\bf (b)} Raw camera image containing the single-shot spectra of the two polarization outputs $S_{1,2n}$, and the unmodulated laser spectrum $S_{0n}$. {\bf (c)} and {\bf (d)}: EO signals on two polarizations channels (after background subtraction and normalization, see Methods). {\bf (e)}: Single-shot input signal retrieved from (c) and (d) using the {\DEOS} phase-diversity-based algorithm (red). Green trace: actual input, obtained using scanned electro-optic sampling. Inset: Fourier spectra of the two terahertz signals. {Note that the classical (i.e., single channel) spectral decoding method would just provide the deformed signals (c) or (d), depending on the channel used. More generally, a classical single-channel method would provide an input signal with good fidelity only if its bandwidth is small compared to the location of the first transfer function zero (dashed lines in the inset of (e)), or if the signal bandwidth is within one of the lobes in the frequency response in Fig.~\ref{fig:fig1}b}}
\label{fig:TDS}
\end{figure*}

%\subsection{Experimental setup}
In order to test the {\DEOS} method experimentally, we first constructed the setup displayed in Figure~\ref{fig:TDS}a. Terahertz pulses are produced by optical rectification of 800~nm millijoule-range laser pulses in a Zinc Telluride (ZnTe) crystal. These terahertz pulses are then analyzed by an EO sampling setup based on our design.

% The stretcher is a classical Treacy compressor. The imaging spectrometer is composed of a reflection grating, a low-cost 1280x1024 pixels monochrome CMOS camera (UI~3240~ML~NIR from IDS GmbH) equipped with a 60~mm objective (Nikkor 60~mm F2.8G ED). We also placed a cylindrical lens with 100~mm focal length just before the 60~mm lens, in order to spread vertically the optical power onto the CMOS camera. The three spots are vertically binned at analysis stage, thus increasing the equivalent full-well capacity (and the signal-to-noise ratio). 

% At each shot $n$, the recorded image (see Figure~\ref{fig:TDS}b) provides three raw data: the spectra on the two polarization channels $S_{1n}(\lambda)$ and $S_{2n}(\lambda)$ (containing the information on the terahertz field), and the spectrum of the unmodulated laser $S_{0n}(\lambda)$. The camera is trigged by the laser, and acquires 10 images per second. For this test we did not try to achieve 1~kHz repetition rate (i.e., the repetition rate of the laser), although this type of upgrade would be straightforward using a state-of-art commercial CMOS or CCD camera (as, e.g., in Ref.~\cite{tikan2018single}).

For test purposes, we also designed the setup so that it is possible to skip or operate the grating stretcher, without changing the setup alignment. We can thus analyze the terahertz pulses using either single-shot "realtime" EO sampling (i.e., using chirped pulses), or using the traditional "equivalent-time" EO sampling by scanning the delay between the femtosecond laser pulses and the terahertz signal. We will thus test our single-shot method by comparing the {\DEOS} results to the corresponding scanned EO signals.

%\subsection{Experimental results}
A typical result is presented in Figure~\ref{fig:TDS}e, using a time window of 20~ps. The reconstructed EO signal and the reference EO signal (i.e., obtained using scanned electro-optic sampling) are found to be extremely similar in shape. The reconstruction is even able to reproduce fine details, as the small oscillations between 5 and 20~ps which are due to the water vapor absorption in air (free-induction decay), and multiple reflections on the Silicon plate. For comparison, the EO signal shapes $Y_{1,2}$ before reconstruction (i.e., corresponding to the previous state-of-art) are very different from the input, as can be seen in Figure~\ref{fig:TDS}c,d.

These results confirm that the {\DEOS} method now enables investigations of terahertz sources, as well as Time-Domain Spectroscopy to be achieved in single-shot, with simultaneously high temporal and spectral resolution. More precisely the time resolution and bandwidth limit of the {\DEOS} method appear -- as theoretically expected -- similar to classical (and non-single-shot) "scanned" electro-optic sampling. 

\subsection*{Note: determination of the reconstruction parameter}

As this type of reconstruction requires the knowledge of the transfer functions $H_{1,2}$, it is important to find a practically convenient approach for determining the parameter $B$. We remarked that $B$ can be determined in a very simple way, by analyzing the recorded data corresponding to the unknown signal. From the reconstructed signal $\phiin^{retr}(t)$, we can simulate the corresponding {\DEOS} signals $\tilde Y_1^{retr}(\Omega)$ and $\tilde Y_2^{retr}(\Omega)$:
\begin{eqnarray}
\tilde Y_1^{retr}(\Omega)=H_1(\Omega)\tilde{\phiin^{retr}}(\Omega)\\
\tilde Y_2^{retr}(\Omega)=H_2(\Omega)\tilde{\phiin^{retr}}(\Omega)
\end{eqnarray}
where $H_{1,2}$, $\tilde Y_{1,2}^{retr}$ and $\tilde{\phiin^{retr}}(\Omega)$ depend on $B$.

\begin{figure}
\begin{center}
\includegraphics[width=8cm]{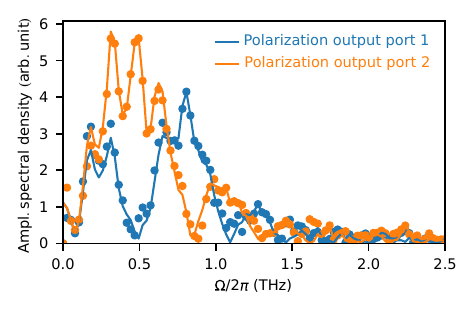}
\end{center}
\caption{Fit providing the reconstruction parameter $B$ from a single-shot recording. Dots: Fourier spectra $|\tilde Y_{1,2}(\Omega)|$ of experimental data before reconstruction. Lines: spectra $|\tilde Y_{1,2}^{retr}(\Omega)|$ computed from the retrieved input. $\tilde Y_{1,2}^{retr}(\Omega)$ are fitted on $\tilde Y_{1,2}(\Omega)$ using $B$ as a free parameter. Note the presence of interleaved zeros, which play a key role in the retrieval approach. Same data and color codes as for Figure~\ref{fig:TDS}c,d.}
\label{fig:fit}
\end{figure}

Then we can perform a least-square fit of {$\tilde{Y_1^{retr}}$ and $\tilde{Y_2^{retr}}$ on $\tilde{Y_1}$ and $\tilde{Y_2}$}, using $B$ as a free parameter. Here, we perform a classical least-square fit using the following definition for the reconstruction error $\epsilon$:
\begin{eqnarray}
\epsilon^2=\int_{-\infty}^{+\infty}d\Omega \left(\left|\tilde Y_1-\tilde Y_1^{retr}\right|^2+\left|\tilde Y_2-\tilde Y_2^{retr}\right|^2\right)
\end{eqnarray}
Besides its use for checking the fit quality, it is interesting to examine the reconstructed $\tilde Y_{1,2}^{retr}$ and raw $\tilde Y_{1,2}$ EO signals (Figure~\ref{fig:fit}). Those curves clearly display the nulls that stem from the zeros of the transfer function $H_{1,2}$ expressed in Eqns.~(\ref{eq:H1},\ref{eq:H2}). 

Remarkably, this result shows that the determination of the fit parameter $B$ does not require specific experiments to be performed. The fit can be made using any unknown input signal -- including the signal to be analyzed -- provided its bandwidth is sufficient.

\subsection*{{Application to measurements of relativistic electron bunch shapes at the European X-ray free-electron laser}}
The {high resolution retrieval} based on phase diversity is
expected to find immediate applications in
{single-pass} Free-Electron Lasers (FELs), such as EuXFEL~\cite{decking2020mhz}. {FELs
  are based on {self-}amplification of
  {stimulated radiation driven} by ultra-short
  relativistic electron bunches with high peak current propagating in
  a periodic magnetic field {(undulators)}. They
  {provide} femtosecond {photon}
  pulses with energies in the millijoule range,
  and wavelengths ranging from the ultraviolet to
  {hard} X-rays depending on the
  facilities. {The emitted photon pulses} depend
  directly on the properties of the electron bunches. Hence the 
  diagnostics of their longitudinal shape has been the subject of an
  intense research the last
  years~\cite{PhysRevLett.88.124801,casalbuoni2008numerical,steffen2009electro,wu2014comparison,steffen2020compact}.

  {A well-known and efficient way to measure the shape of a high energy electron bunch consists of probing its Coulomb field, at a short distance $d$. The fundamental time-resolution limit of such a measurement is directly determined by the Coulomb field distribution of a single electron in the laboratory frame (see, e.g., Ref.~\cite{Coulomb_field_Jackson}). The resulting resolution limit is of the order of:~\cite{EOS_acc_before_FELIX,yan2000subpicosecond}
\begin{equation}
\tau_{R}^{Coulomb}\approx \frac{2d}{\gamma c},\label{eq:resolution_coulomb}
\end{equation}
with $\gamma$ the Lorentz contraction factor, and $c$ the speed of light. For energies in the GeV range (i.e., $\gamma$ of the order of several thousands for electrons) and a distance $d$ of few millimeters, this time resolution limit $\tau_R^{Coulomb}$ lies in the few tens of fs or even below.
}

{As a result, in high energy machines, the main resolution limit is actually set by the capabilities of state-of-art single-shot photonic measurement systems.} In particular, Eq.~(\ref{eq:resolution_limit}) strongly
  hampered the application range of classical spectrally-decoded electro-optic sampling to values well above the limit set by Eq.~(\ref{eq:resolution_coulomb}). The
  situation is also complicated by the recent trend forward repetition rates in the megahertz-range, in
  X-ray FELs {based on superconducting technology} such as FLASH~\cite{Ackermann:2007aa}, EuXFEL~\cite{decking2020mhz}, and the LCLS-II~\cite{Galayda2018} and SHINE~\cite{shine} projects.}

\begin{figure}[htbp!]
\includegraphics[width=8cm]{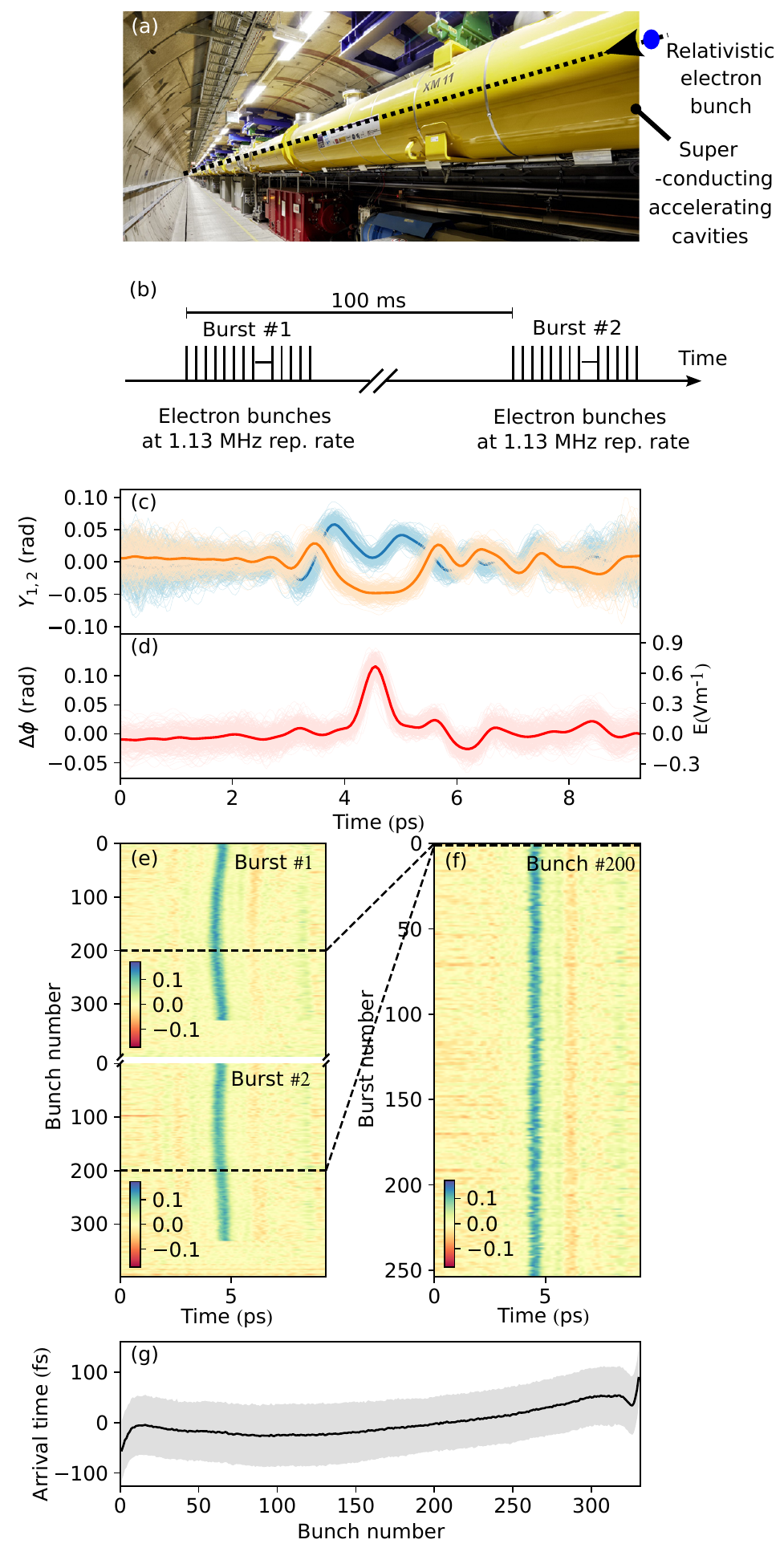}
\caption{Electron bunch shapes recorded at the European X-ray Free-Electron Laser (EuXFEL). {\bf (a)} {Picture from inside the 3km-long accelerator tunnel}. Our {\DEOS} setup (see Fig.~\ref{fig:XFEL_setup}) is placed just upstream of the picture, after the first bunch compressor. {\bf (b)} timing of the electron bunches in the conditions of the experiment. {\bf (c)} Electro-optic signals $Y_{1,2}$ of a single bunch before reconstruction. {\bf (d)} Reconstructed electric field. Shaded areas: superposition of single-shot curves, color curves: average over $255$ bursts. {\bf (e)} Electro-optic signal of two bursts (i.e., $800$ electron bunches in total). {\bf (f)} Shape of one bunch (with bunch number $200$ within the burst) versus burst number. {\bf (g)} Arrival time versus bunch number. Shaded areas:  RMS arrival time fluctuations, color curve: average over $255$ bursts. The EO data are low-pass filtered to $2.5$~THz.}
\label{fig:XFEL}
\end{figure}

%\subsection{Experimental setup}
 In order to test our {\DEOS} reconstruction technique in this context, we
  realized a proof-of-principle phase-diversity setup at the EuXFEL (Figure~\ref{fig:XFEL}a), which is the first hard X-ray
  {FEL} operating at megahertz repetition
  rate~\cite{decking2020mhz} (see also Refs.~\cite{Gisriel2019,Pandey2020} for examples of applications). 
In the present case, electron bunches are generated at 1.3~MHz rate in 600~$\mu$s long bursts, every 100~ms (Figure~\ref{fig:XFEL}b). 

  The setup, based on the single EO channel detection~\cite{steffen2020compact}, is depicted schematically in
  Figure~\ref{fig:XFEL_setup}. The EO setup, {destined to record the Coulomb field versus time}, is located just after the
  second bunch compressor, where the electron bunches
  {have a beam energy of 700~MeV and a typical RMS
    duration in the 200~fs range at a charge of 250 pC}. {The distance $d$ between the laser and the electron bunch is of the order of 5~mm which corresponds to a time resolution limit $\tau_R^{Coulomb}$ in the $30$~fs range for the measurement of the Coulomb field}. Note that, in contrast to the previous experiment, where real single-shot operation was achieved, we had here to record each polarization channel {\it successively}, due to the lack of a second {MHz line rate spectrometer} (see Methods). Hence the present proof-of-principle experiment is not single-shot for the moment. However, a relatively obvious upgrade is planned in order to achieve simultaneous recordings of the two channels.

%\subsection{Results}
Results are displayed in
  Figure~\ref{fig:XFEL}c,d. Figure~\ref{fig:XFEL}c
  {depicts} the raw EO signals (after background
  subtraction and normalization), and Figure~\ref{fig:XFEL}d
  represents the reconstructed electric field. We observe a main peak,
  {for which the duration of $218$~fs (RMS) is in very good agreement with the design} value of the electron bunch duration at this location
  {and considerably shorter than measured with
    conventional spectrally decoded EO sampling at the same
    setup~\cite{steffen2020compact}}. The smaller negative peak can be
  attributed to a wakefield following the electron bunch, but the
  precise interpretation will be subject of further
  investigations. Typical examples are
  displayed in Figures~\ref{fig:XFEL}e,f.  This measurement system can now be used to perform
  bunch-by-bunch high resolution measurements of the bunch shape which is crucial information for the control of the bunch compression process. 

{As previous spectrally decoded EO systems, {\DEOS} also simultaneously measures the arrival time of the electron bunch, which is also a crucial parameter for users of the generated X-rays. The resolution of {\DEOS} for this measurement is expected to be similar to standard EO sampling. In the present case, the arrival time jitter (Fig.~\ref{fig:XFEL}g) is measured to be $58$~fs over a bunch train, which is much lower than the bunch duration ($218$~fs~RMS), and consistent with Ref.~\cite{steffen2020compact}. 

% Note that the 58~fs measured time arrival jitter is much lower than previously measured values on the Free-Electron Laser X-ray pulses (308~fs in Ref.~\cite{kirkwood2019initial}). This motivates further studies aiming at mastering the X-ray FEL pulse arrival time.

{
From the hardware point of view, it is important to note that only few {key} modifications of our initial EO system~\cite{steffen2020compact} were needed: ensuring a proper (non-standard) orientation of the GaP crystal in the vacuum chamber, and {simultaneous detection of} the two EO output channels. We thus think that the implementation of a double EO output channel readout will permit to relatively easily adopt the phase diversity-based retrieval method in existing or planned EO diagnostics {at FELs or other accelerators}. 
}

\begin{figure}[htbp!]
\begin{center}
\includegraphics[width=8cm]{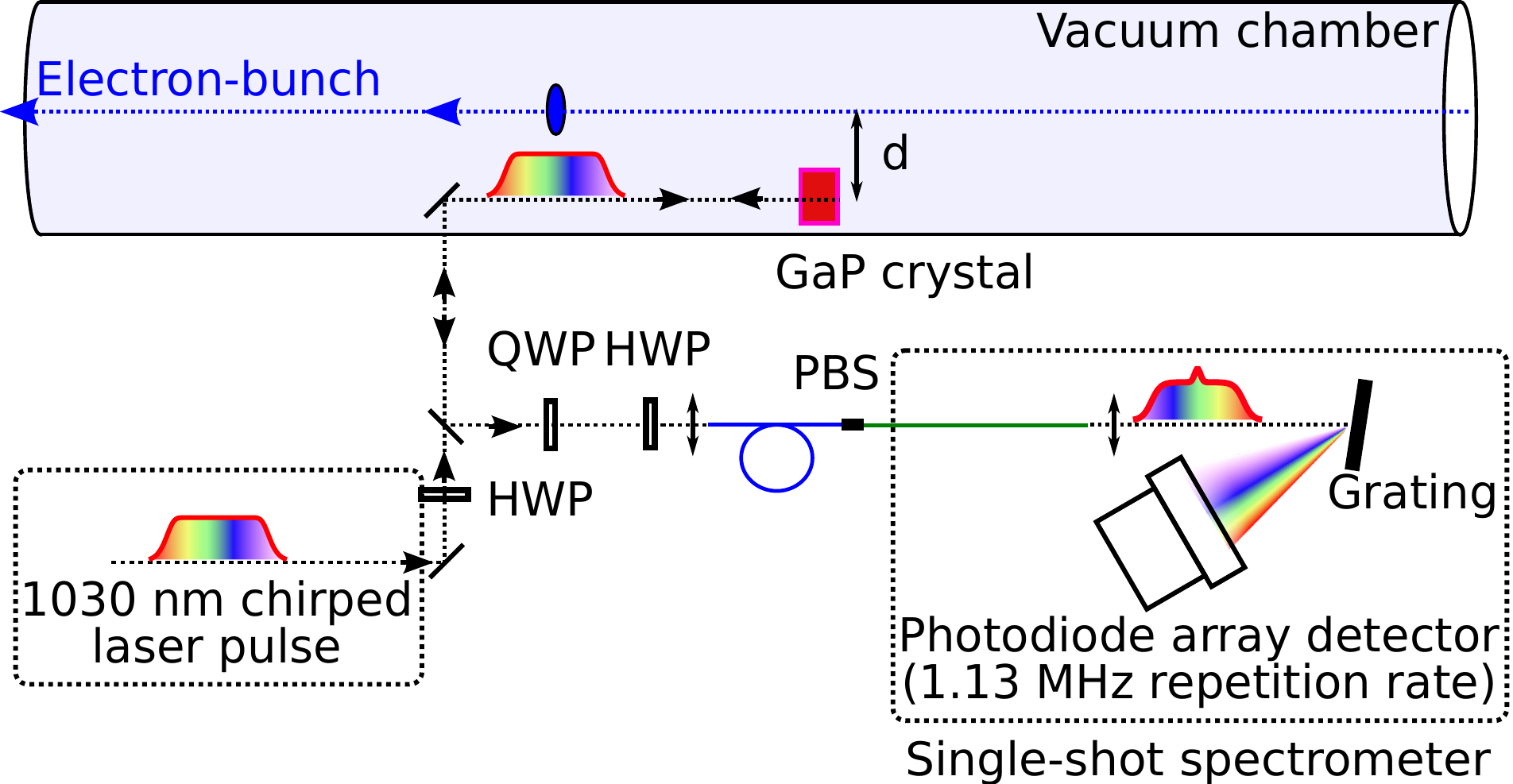}
\end{center}
\caption{{{\DEOS} measurement of the Coulomb field created by relativistic electron bunches at the European X-ray Free-Electron Laser (Eu-XFEL).}{A laser beam probes the electric field at a distance $d=5$~mm from the electron beam.} HWP: half wave plate, QWP, quarter wave plate, PBS, fiber-based polarizing beam-splitter. Blue lines indicate polarization-maintaining fibers (PM980) and green lines indicate single-mode fibers (HI1060). The probe laser is reflected on a Gallium Phosphide (GaP) crystal back side. The spectrum readout is performed using a KALYPSO linear array camera operating at 1.13~MHz {line} rate~\cite{Gerth19,rota2018kalypso}. Details of laser transport and KALYPSO focusing optics are not shown -- see Ref.~\cite{steffen2020compact} for further details.}
\label{fig:XFEL_setup}
\end{figure}

\section*{Discussion}

It is important to note that {\DEOS} is fundamentally different from previous techniques aiming at retrieving the input field using numerical analysis of the EO data (deconvolution~\cite{wu2018electro}, and holography-inspired reconstructions~\cite{yellampalle2005algorithm}). The presence of zeros (nulls) in the transfer functions {$H_1(\Omega)$ and $H_2(\Omega)$} limits the success of reconstruction {methods that do not use the phase diversity technique} to relatively narrowband signals, or short analysis windows, if there is no a priori information on properties of the solution. {More precisely, deconvolutions or reconstructions from a single channel is ill-posed when the signal spectrum $\Omega_{max}$ exceeds the angular frequency $\Omega_0$ of the first null: $\Omega_0=\sqrt{C\pi/2}$ or $\sqrt{3C\pi/2}$ (depending on the channel used) for the Figure~\ref{fig:fig1} setup. The limit is $\Omega_0=\sqrt{C\pi}$ for the case of the more classical situation for which the [-110] axis is parallel to the terahertz field (see Section~V of Supplementary Material for the corresponding transfer functions).} 

{In contrast, the DEOS strategy allows a numerical reconstruction to be possible on a frequency range that extends well above the nulls of $H_1$ and $H_2$. After applying the MRC algorithm, the resulting transfer function is flat -- by design -- as shown in Eq.~(\ref{eq:HMRC}). In consequence, DEOS enhances the effective bandwidth over which a THz signal can be reliably recorded without distortion.
%  {The theoretical transfer function of DEOS, defined from the input THz-induced crystal phase-shift $\Delta\tilde\phi(\Omega)$ to the reconstructed signal $\Delta\tilde\phi_{in}^{retr}(\Omega)$ is -- by design -- perfectly flat. This can be easily verified by injecting the definitions of the single channel transfer functions [Eqns.~(\ref{eq:H1},\ref{eq:H2})] in the MRC reconstruction formula~[Eq.~\ref{eq:MRC})]. One easily finds:}
% \begin{eqnarray}
% &&\Delta\tilde\phi_{in}^{retr}(\Omega)=H_{MRC}(\Omega)\Delta\tilde\phi_{in}(\Omega),\\
% &&\text{with}\; H_{MRC}(\Omega)=1,
% \end{eqnarray}
% {or, after an inverse Fourier transform:}
%  \begin{eqnarray}
% \Delta\phi_{in}^{retr}(t)=\Delta\phi_{in}(t).
% \end{eqnarray}
% {This implies that any sharp input temporal feature $\Delta\phi_{in}(t)$ with bandwidth $\Omega_{max}$ can be theoretically resolved and retrieved without distortion, provided the corresponding input data $\tilde Y_1(\Omega)$ and $Y_2(\Omega)$ can be obtained with reasonable signal-to-noise ratio, up to $\Omega_{max}$.}
In practice, the ultimate temporal resolution $\tau_R^{DEOS}$ is hence conjectured to be limited by either (1) the laser pulse bandwidth $\Delta\nu_L$, or (2) the bandwidth $\Delta\nu_{crystal}$ of the EO crystal, whichever is slower.} {Bandwidth limitations by the electro-optic crystal have been previously extensively studied, and will not be discussed here. The main limitations comes from the phase-matching (i.e., the crystal thickness) and phonon absorption lines, which are at 5.3, 8.0 and 11~THz for classical Zincblende crystals ZnTe, GaAs and GaP~\cite{wu2014comparison}, respectively. Note that the crystal speed limitation affects the response of the Pockels crystal phase shift $\Delta\phi(t)$ with respect to the input electric field $E_{THz}(t)$.}

{DEOS specifically enhances the bandwidth and time resolution of the next part of the system, i.e., from the crystal's output phase shift $\Delta\phi_{in}(t)$ to the final (retrieved) EO signal $\Delta\phi_{in}^{retr}(t)$. It is hence natural to introduce a figure of merit that relates the available laser bandwidth (which is a fundamental limitation to the measurement) and the corresponding resolution limitation:
\begin{eqnarray}
\tau_R^{DEOS} = \eta_L\frac{1}{\Delta\nu_L}. \label{eq:tau_limit_laser}
\end{eqnarray}
In this Fourier reciprocal relation, the introduced figure of merit $\eta_{L}$ has the meaning of a time-bandwidth product. Systematic studies of the resolution versus available laser bandwidth will require specific research, for instance in order to find an analytic bound to $\eta_L$. However, the resolution $\tau_R^{DEOS}$ can already be easily computed numerically in simple cases. As an example, in Table~\ref{table:resolution} and in Section~VIII of the Supplementary Material, we present such a preliminary study in the case of the Eu-XFEL experiment. {In this case (and with all widths being defined FWHM) we find that the figure of merit $\eta_L$ is slightly below unity, and the corresponding time resolution limit (for measuring a Gaussian THz pulse) is approximately 2.2 times the laser pulse duration (see Table \ref{table:resolution}).}}
\begin{table*}[htbp]
\begin{center}
\begin{tabular}{ |c|c|c| } 
\hline
Laser pulse bandwidth& {$\Delta\nu_L=11.5$~THz FWHM}\\
\hline
Corresponding compressed laser pulse duration& $\tau_L=39~fs$ FWHM\\ 
\hline
Laser chirped pulse duration at the Pockels crystal& $\tau_w=10$~ps FWHM\\ 
\hline
Temporal resolution of the DEOS measurement(computed) & {$\tau_R^{DEOS}=84.5$~fs FWHM} \\
\hline
Figure of merit & {$\eta_L=\Delta\nu_L\times \tau_R^{DEOS} =0.94$} \\
\hline
\end{tabular}
\end{center}
\caption{{Example of expected time resolution limitation (numerical simulation) due to the finite laser bandwidth (i.e., an infinitely fast crystal is assumed). Typical parameters corresponding to the Eu-XFEL experiment are used, i.e., a laser with 1040~nm wavelength and 40~mn bandwidth FWHM. The THz and laser pulses are taken Gaussian and we consider a chirped pulse duration $\tau_w=10$~ps. See Section~VIII of the Supplementary Material for details, including the dependence of the time resolution $\tau_R^{DEOS}$ with respect to the chirped laser probe duration $\tau_w$.}}
\label{table:resolution}
\end{table*}

{In this work, it is interesting to note that, even though DEOS could straighforwardly beat the historical limit set by Eq.~(\ref{eq:resolution_limit}), we did not have the possibility to produce THz signals presenting features short enough to test the "new" ultimate limit set by the laser bandwidth [Eq.~(\ref{eq:tau_limit_laser})], nor the crystal bandwidth limitation. Even in the FEL experiment, the shortest features we could resolve were the bunch shape itself, which has an RMS duration in the 200~fs range, whereas the resolution limit set by the laser bandwidth is numerically predicted (see Table~\ref{table:resolution}) to be less than 100~fs~FWHM.}

 {Besides, another equally important property of DEOS lies in the possibility to considerably extend the duration of the analysis window $\tau_w$, while keeping the bandwidth of analysis (or time resolution) constant. This is important in particular for time-domain spectrometers (as the one presented in Fig.~\ref{fig:TDS}), since the spectral resolution is directly determined by $1/\tau_w$. A long recording window is also crucial in studies of the dynamics of various sources, as THz Free-Electron lasers~\cite{hafez2018extremely}, Coherent Synchrotron Radiation sources~\cite{roussel2015.EOS}, and quantum cascade lasers~\cite{cappelli2019retrieval}. The achievable analysis window $\tau_w$ of DEOS is no longer limited by Eq.~(\ref{eq:resolution_limit}), and this represents a considerable advantage with respect to standard electro-optic sampling. While this suggests that the window of analysis $\tau_w$, can be chosen arbitrarily large,} {component limitations will restrict its achievable range. The most obvious one is the number of pixels $N_{cam}$ of the camera that resolves the optical spectrum (1280 and 256 for the first and second experiments of this article). This sets an upper limit to the "effective number of points" that can be recorded to
\begin{equation}
N_{eff}=\frac{\tau_w}{\tau_R}<N_{cam}.
\end{equation}
Technically, the effective number of points will be also affected by the well-know spectrometer resolution parameters, in particular the number of grating lines that diffracts the laser beam, and aberrations. This will not be discussed here as this does not represent a fundamental limit, and is a well documented topic. However, when needed, care will have to be taken in the spectrometer choice for reaching an effective number of point $N_{eff}$ close to the camera resolution.}
{If applications require an effective number of points $N_{eff}$ beyond camera capabilities, different ways may be possible for recording the spectrum.} {A foreseen alternative} is the so-called {Dispersive Fourier transform}~\cite{jalali_nature_photonics_review} ,
%~\cite{kelkar1999time}\cite{fard2013photonic},
 which may be advantageous from this point of view. This would turn the present recording system to a photonic time-stretch data acquisition system~\cite{time_stretch_first_bhushan1998time}\cite{time_stretch_transfer_function_han_2003}.
 %\cite{fard2013photonic}.

{Future work aiming at reaching the highest bandwidths and/or longest record duration will probably require specific studies, in particular for managing higher order dispersion. Indeed, DEOS theory has been established assuming quadratic probe laser chirp, and unwanted third order dispersion will lead to loss of quality in the reconstruction. It is important to note that the present method implies a fit (see Figure~\ref{fig:fit}) that also provides an objective measurement of the quality of the model-experiment correspondence. Specific works will be needed when this fit will present strong errors, and this will be expected in extreme cases where strong third order dispersion is present and the transfer functions $H_{1,2}$ present many lobes over the bandwidth of the THz signal. These particular cases will reduce the bandwidth of the measurement (to the range allowing a reasonably good fit). Therefore, we think that an important research direction will consist of extending the\textbf{} the DEOS theory in order to take third-order dispersion into account in the reconstruction.
}

{Finally, interesting future directions also concern the quest for the highest possible sensitivity for single-shot {\DEOS}. In this respect, note that the Signal-to-Noise Ratio (SNR) depends on the location inside the laser spectrum, and constant SNR should be achievable by flattennig the probe laser spectrum using Fiber Bragg Gratings-based or programmable optical filters (such as the so-called Waveshaper). Furthermore, {\DEOS} is compatible with the recent SNR-enhancement strategies which has been developed for electro-optic sampling~\cite{szwaj2016high}.}

% It is also important to note that the MRC algorithm considered here is only one possibility, and that other more complex ones should also be considered for {\DEOS}. In particular, the MRC algorithm relies on transfer functions and is thus valid only for small phase modulations in the crystal. In conditions of high electric fields, the formal similarity with the photonic time-stretch ADC will enable back-propagation methods to be efficient~\cite{stigwall2007signal,gupta2007demonstration}.

%\section*{Conclusion}
{In conclusion, we present a novel conceptual framework for spectrally decoded electro-optic sampling, that
 solves the "temporal resolution problem" open about 20~years ago for terahertz recorders. Technically, the key lies in the derivation of the transfer function which -- in turn -- allows well-posed reconstruction algorithms to be possible.} In practice, the resulting {\DEOS} design opens the way to terahertz digitizers whose recording length and resolutions is only limited by the probe crystal speed or the laser uncompressed pulse duration. On the short term, one foreseen application of {\DEOS} concerns a revisit of high repetition-rate Time-Domain Spectroscopy (TDS), up to megahertz rates, and consider monitoring irreversible physical and chemical processes. Another foressen application is totally opposite, and concerns the applications of very low repetition rate terahertz sources, for which scanning strategies are impractical. This concerns nonlinear TDS using accelerator-based~\cite{di2018coherent,hafez2018extremely,pan2019photon}, as well as table-top-based high power terahertz sources~\cite{jolly2019spectral}. Future work will include a systematic experimental study of the new limits set by the method in terms of ultimate resolution, time-window and repetition rate. In this respect, an important direction will consist in combining the present {\DEOS} terahertz recording method with photonic time-stretch~\cite{time_stretch_first_bhushan1998time,roussel2015.EOS,evain2017direct}, as this should theoretically allow repetition rate to reach the hundred of megahertz range.

% Future directions includes investigations of rapidly evolving reactions in chemistry and biology~\cite{hafez2018extremely}, as well as nonlinear physics (such as nonlinear spectroscopy) which use low repetition rate high power terahertz sources. From the industrial point of view, single-shot terahertz recorders would also introduce a disruptive competition with respect to latest fast scanning terahertz spectrometers, such as Asynchronous Optical Sampling (ASOPS) and Electronically Controlled Optical Sampling (ECOPS)~\cite{bartels2007ultrafast,kim2010high}, in the quest for non-destructive and high throughput inspection for, e.g., automotive and drug industries.

\section*{Materials and Methods}
\subsection*{Theoretical details}
Proofs are given in the Supplementary Material, and {we recall only main results here}. The Pockels-induced phase-shift $\phiin(t)$ is related to the terahertz electric field by $\phiin(t)=\factor E(t)$ [see Eqns.~(\ref{eq:H1},\ref{eq:H2})]. The factor $\factor$ depends on the relative orientations of the crystal axes and on the polarizations of the laser and terahertz field. For the phase-diversity scheme of Figure~\ref{fig:fig1} and the setups considered in the main text: 
\begin{eqnarray}
\phiin(t)&=\factor E(t)\\
\text{with}\;\factor&=\frac{\pi d}{\lambda}n_0^3r_{41}
\end{eqnarray}
{where $n_0$ is the refractive index at vanishing electric field, $d$ the thickness of the crystal and $r_{41}$ is its electro-optic coefficient}. $\lambda$ is the laser wavelength in vacuum and $E(t)$ the electric field inside the crystal. See Supplementary Material Section VIII for the values used in the numerical simulations. 
Note that the value of $\factor$ is two times smaller than for the usual approach used in classical balanced detection (see Eqns~(30) and (49) of the Supplementary Material).

\subsection*{Experimental recording system for the table-top experiment (Figure~\ref{fig:TDS})}
The laser pulses are delivered by an amplified Sapphire-Titanium laser (Coherent Astrella)  with 40 fs duration and 7~mJ output, from which~3 mJ are extracted for this experiment. The emission and detection ZnTe crystals are 110-cut, with 1~mm thickness, and {the component orientations are displayed in Table~\ref{table:orientations}}. The stretcher is a classical Treacy compressor. The Silicon filter (280 $\mu$m thick, at normal incidence) is destined to rejected the 800~nm laser light. The imaging spectrometer is composed of a reflection grating  {(Thorlabs, 1200 lines/mm, blazed at 750~nm)}, a low-cost 1280x1024 pixels monochrome CMOS camera (UI~3240~ML~NIR from IDS GmbH) equipped with a 60~mm objective (Nikkor 60~mm F2.8G ED). We also place a cylindrical lens with 100~mm focal length just before the 60~mm lens, in order to spread vertically the optical power onto the CMOS camera. The three spots are vertically binned at analysis stage, thus increasing the equivalent full-well capacity (and the signal-to-noise ratio). 

At each shot $n$, the recorded image (see Figure~\ref{fig:TDS}b) provides three raw data: the spectra on the two polarization channels $S_{1n}(\lambda)$ and $S_{2n}(\lambda)$ (containing the information on the terahertz field), and the spectrum of the unmodulated laser $S_{0n}(\lambda)$. The camera is trigged by the laser, and acquires 10 images per second. {In order to ensure single-shot measurements, we systematically chose a camera exposure time that was shorter than the laser repetition period of 1 ms (the exposure time was 25~$\mu$s for the data presented here).} 

For this test we did not try to achieve 1~kHz repetition rate (i.e., the repetition rate of the laser), although this type of upgrade would be straightforward using a state-of-art commercial CMOS or CCD camera (as, e.g., in Ref.~\cite{tikan2018single}).

\subsection*{Data anaysis in the table-top experiment} Raw experimental data consist of single-shot camera images (Fig.~\ref{fig:TDS}b). At each shot $n$, we extract the three spots: the upper and lower spot corresponding to the EO signals along the two polarizations 1 and 2, and the central spot, which corresponds to the reference  laser spectrum without electric field. This latter is used for correcting the shot-to-shot fluctuations of the laser. The three corresponding spectrum are extracted and provide 1-dimensional arrays $S_{1n}[i]$, $S_{2n}[i]$ and $S_{0n}[i]$ (with $i=0..1279$ corresponding to the horizontal camera row index, and physisally to the wavelength $\lambda$). In addition we also record beforehand the same data in absence of electric field, which provide the three arrays $S_1^{ref}[i]$ and $S_2^{ref}[i]$ and $S_0^{ref}[i]$.

For obtaining the EO signals $Y_{1n}$ and $Y_{2n}$ at each shot $n$ (displayed in Fig.~\ref{fig:fig1}b) in the following way. We first apply the background subtraction and normalization:
\begin{eqnarray}
Y_{1n}^{raw}=\frac{S_{1n}}{\sigma_1 S_{0n}}-1\label{eq:dataprocessing800nm_1}\\
Y_{2n}^{raw}=\frac{S_{2n}}{\sigma_2 S_{0n}}-1\label{eq:dataprocessing800nm_2}
\end{eqnarray}
where $\sigma_{1,2}=\frac{S_{1,2}^{ref}}{S_0^{ref}}$.

Afterwards, we discard the low-SNR data which correspond to the low intensity wings of the laser spectrum, by multiplying the data by a window function $W$.
\begin{eqnarray}
Y_{1n}[i]=W[i]Y_{2n}^{raw}[i].\label{eq:window}
\end{eqnarray}
In the experimental results of Figure~\ref{fig:TDS}, we used a simple square window, with $W[i]=1$ for pixels rows $i=300$ to 1000, and $W[i]=0$ elsewhere.

Note that the first part of the preprocessing (\ref{eq:dataprocessing800nm_1},\ref{eq:dataprocessing800nm_2}) involves a measurement of the laser spectrum $S_{0n}$, so that shot-to-shot laser spectrum fluctuations are canceled out. In other words, $Y_{1,2n}$ do not depend on the shape of $S_{0n}$. This can be shown easily, if we assume a fixed relation between the three spectra: $S_{0,1,2n}(\lambda)=f_{0,1,2}(\lambda)S_{Ln}(\lambda)$, with $S_{Ln}(\lambda)$ the laser spectrum at shot $n$, and assume that $f_{0,1,2}(\lambda)$ are functions that do not change with time (i.e., they only are determined by the optics transmisions, and adjustment). This preprocessing should theoretically allow the measurements to reach the shot-noise limit.

The obtained $Y_{1,2n}$ are then used in the MRC retrieval algorithm. 
% We first apply a window in order to select the part of $Y_{1,2n}(t)$ for which the laser spectrum intensity is well above the detection noise (in practice we select the part where the signal is larger than 10$\%$ of the maximum).
 Then:
\begin{itemize}
\item We apply a Fast Fourier Transform (FFT) to the windowed $Y_{1,2n}(t)$ data.
\item Apply the MRC formula displayed in Eq.~(\ref{eq:MRC}).
\item Finally apply and inverse FFT for obtaining the reconstructed data $\Delta\phi_{in}^{retr}(t)$ and $E_{in}^{retr}(t)$.
\end{itemize}

\subsection*{Other option for the reconstruction}
Note that the data processing may be performed in another order. Instead of first applying the background subtraction and normalization~(\ref{eq:dataprocessing800nm_1},\ref{eq:dataprocessing800nm_2}) before the MRC algorithm, we can also first apply  the background subtraction, then MRC, and eventually the normalization. The results are extremely similar for all the experiments presented in this article (see Supplementary Material Section~VIII for a reconstruction using this order).  

\subsection*{Experimental setup at EuXFEL (Figure~\ref{fig:XFEL_setup})}

The probe pulses are devivered by an amplified Ytterbium fiber laser operating at 1030~nm. The EO effect is achieved in a Gallium Phosphide (GaP) crystal with 2~mm thickness, which is
placed inside the vacuum chamber of the accelerator, near the
electron beam~\cite{steffen2020compact}. The {current} setup permits to
detect only one EO channel output at a time, because one readout camera is available at the time.

The spectra are recorded in single-shot, using a grating spectrometer based on the KALYPSO
fast linear array {detector~\cite{rota2018kalypso,Gerth19} operated at a line
rate of 1.13~MHz}. A first series of electron bunch trains for one
polarization is recorded and then the last half-wave plate  before the
polarizing beam splitter is rotated by $\pi/8$ to record the
complementary polarization. Hence, the two EO channel outputs can be
used to reconstruct the individual electron bunch shapes within the
burst using the phase-diversity technique. Note that a straightforward upgrade is also planned, with the aim to achieve simultaneous recording of both polarisations, and realize true single-shot operation.}

%This allowed to achieve up to 1.13~MHz measurement rate (i.e., $1.13\times 10^6$ EO traces per second). 

% This limited the acquisition rate to half of the Eu-XFEL repetition rate (2.26~MHz) achieved during this experiment, i.e., every other bunch was recorded in this study. This is however not a strong limitations, as the recent upgrade of KALYPSO is able to operate at 4.52~MHz and 1024 pixels.

\subsection*{Data analysis in the EuXFEL experiment}
{With each burst, we record single-shot spectra on one of the two polarization directions $S_{1n}(\lambda)$, $S_{2n}(\lambda)$ and additionally unmodulated spectra with laser and no electron bunch $S_{1,2}^{no\;bunch}$, and the spectra without laser $S_{1,2}^{dark}$.}
 
{However as we do not record the unmodulated laser spectra $S_{0n}(\lambda)$ of the same laser pulses in the XFEL experiment, we cannot compensate for the shot-to-shot fluctuations of the laser faster than about 2~kHz. The data analysis is then similar to the 800 nm experiment case. 
We thus define the EO signals (before applying the MRC reconstruction) as:}
{ \begin{eqnarray}
Y_{1n}^{raw}=\frac{S_{1n}-S_1^{dark}}{S_{1}^{no\;bunch}-S_1^{dark}}-1\\
Y_{2n}^{raw}=\frac{S_{2n}-S_2^{dark}}{S_{2}^{no\;bunch}-S_2^{dark}}-1,
\end{eqnarray}}
And as before, we multiply the data by a window function [see Eq.~(\ref{eq:window})]:
\begin{eqnarray}
Y_{1,2n}[i]=W[i]Y_{1,2n}^{raw}[i]
\end{eqnarray}
We chose here a Tukey function for $W[i]$, which falls to zero at the points where the laser spectrum is at 5\% of its maximum, and with a shape parameter equal to 0.25. Then the MRC algorithm in exactly the same way than for the previously described table-top experiment.

\section*{Acknowledgements}
The PhLAM team was supported by the following funds: Ministry of Higher Education and Research, Nord-Pas
de Calais Regional Council and European Regional Development Fund (ERDF) through
 the Contrat de Plan
\'Etat-R\'egion (CPER photonics for society), LABEX CEMPI project (ANR-11-LABX-0
007), and the ANR-DFG ULTRASYNC project (ANR-19-CE30-0031). ER was supported by the METEOR CNRS MOMENTUM grant. BS and CG acknowledge support from DESY, a member of the Helmholtz Association HGF. BJ (UCLA) was supported by the Office of Naval Research (ONR) Multi-disciplinary University Research Initiatives (MURI) program on Optical Computing Award Number N00014-14-1-0505. This work also used the PhLAM femtosecond laser facility in 2016 and 2017. The authors would like to thank Marc Le Parquier and Nunzia Savoia (PhLAM) for their 
work on the operation of the Titanium-Sapphire laser. 

\section*{Conflict of interests}
The authors declare that they have no conflict of interest

\section*{Contributions}
 The PhLAM team realized the theoretical and numerical investigations: establishment of transfer function versus crystal arrangement, analysis algorithm based on MRC, and simulation codes. These investigations have been based on the concepts established by UCLA (BJ) on transfer functions, phase diversity and MRC, in the framework of photonic time-stretch. Theoretical and numerical calculations have been performed by SB and ER. Table-top experiments have been designed and realized and operated by CS, ER, CE and SB. The  EuXFEL {\DEOS} system has been designed and realized by BS and CG. The setup modifications for phase-diversity studies at EuXFEL have been designed by the PhLAM and DESY teams, and realized by BS. Data analysis has been performed by ER and SB (table-top experiments), and ER, BS, CS, and SB (FEL experiments). All authors participated to the manuscript redaction.

\section*{Supplementary information} accompanies the manuscript on the Light: Sciences and Applications website \url{https://www.nature.com/articles/s41377-021-00696-2}.

\section*{References}

% \section*{Data availability}
% The data that support the plots within this paper and other findings of this study are available from the corresponding author upon reasonable request.

% \section{References}
\bibliographystyle{naturemag}
%\bibliography{eos_phase_diversity1}

\end{document}